\documentclass{emulateapj}
\shorttitle{Wave Decay in MHD Turbulence}
\begin{document}

\title{Wave Decay in MHD Turbulence}
\author{Andrey Beresnyak and Alex Lazarian}
\email{andrey, lazarian@astro.wisc.edu}
\affil{University of Wisconsin-Madison, Dept. of Astronomy}
\journalinfo{The Astrophysical Journal, 678:961-???, 2008 May ??
\hfil\break\copyright\ 2008. The American Astronomical Society. All rights reserved.}

\begin{abstract}
We present a model
for nonlinear decay of the weak wave in three-dimensional
incompressible magnetohydrodynamic (MHD) turbulence.
We show that the decay rate is different for parallel
and perpendicular waves.
We provide a general formula
for arbitrarily directed waves and discuss particular limiting
cases known in the literature.
We test our predictions with direct numerical
simulations of wave decay in three-dimensional MHD turbulence,
and discuss the influence of turbulent damping
on the development of linear instabilities in the interstellar
medium and on other important astrophysical processes.
\end{abstract}


\section{Introduction}
The interstellar medium (ISM) is turbulent on scales ranging from AUs
to kpc (see Armstrong et al. 1995, Elmegreen \& Scalo 2004) with an
embedded magnetic field that influences almost all of its properties.
Turbulence is essential for understanding the ISM, the
intracluster medium (ICM), the Earth's magnetosphere,
solar wind, accretion disks, etc.

The literature on astrophysical turbulence and its applications is
vast (see Biskamp~2003, McKee \& Ostriker 2007 and references
therein). The concepts of waves and eddies are both used to describe
magnetized turbulence. It is accepted that weak Alfv\'enic turbulence
consists of waves, while for strong turbulence wave-eddy
dualism is important (Goldreich \& Sridhar 1995, henceforth GS95).
In strong Alfv\'enic turbulence, nonlinear interactions damp the
energy of wave-eddies over approximately one wave period.
However, there are quite a few sources which exist in the turbulent
environment that emit waves with different degrees
of monochromaticity. These waves propagate in
the turbulent medium while being modified by the medium.
The evolution of such waves is of considerable astrophysical interest
and has been addressed in the literature on numerous occasions.
For instance, collisional and collisionless damping of fast
modes in a turbulent
medium was shown to differ considerably from the case when waves
propagate along laminar magnetic field lines, with important
consequences for cosmic ray propagation (Yan \& Lazarian 2004). The
latter, however, is the linear damping of waves, while the focus of this
paper is the nonlinear interaction of Alfv\'en waves with turbulence.

Since we are addressing the subtle issue of nonlinear damping of Alfv\'en
waves by
surrounding strongly nonlinearly damped Alfv\'enic eddies, we feel that a
discussion of the properties of MHD turbulence is due.
It is well known that magnetohydrodynamics (MHD)
describes the dynamics of a highly conducting fluid with magnetic field.
Highly disordered quasi-stochastic flow, usually called turbulence, is
generally observed in settings when the Reynolds number, the ratio of
the inertial force to the viscous force, is large. When conductivity of the
fluid is very low, it can be described rather well by the Navier-Stokes
equations, which are, basically, the MHD equations without magnetic
field. If the conductivity, however, is high, the magnetic field can
no longer be ignored. In this case, the {\it magnetic Reynolds number} $Re_m$, or the ratio of the fluid velocity to the magnetic diffusion velocity, is
large. In fact, in the turbulent flow of a conductive
fluid there is a mechanism called dynamo, which increases total
magnetic energy, making it more dynamically important over time (see,
e.g., Vishniac et al, 2003).  It is no wonder that in the
astrophysical environment, where both Reynolds and magnetic Reynolds
numbers are high, magnetic fields are always present and always dynamically
important.

The paradoxical nature of turbulence is that even though the dissipation
parameters (e.g. viscosity and magnetic diffusivity) tend to zero,
the dissipation rates stay pretty much the same.
The general explanation for this, as proposed by Kolmogorov (1941),
is that the energy is carried from large scales through many
intermediate scales down to small dissipation scales.
This property is shared by turbulence in magnetized fluids.
Much of the progress in understanding turbulence comes from
so-called Kolmogorov models which assume a cascade of a
conserved quantity, such as energy, down-scale. Once the functional
dependence of the cascade decay rate, local in k-space,
is known, we roughly know the spectrum of the conserved quantity
(normally energy, but sometimes other quantities, such as the number of quasiparticles,
enstrophy, etc.)\footnote{The Kolmogorov-type models do not necessarily
give $-5/3$ spectrum, which is the result specific for the incompressible
hydro turbulence. For example, so-called Langmuir turbulence has the cascade
direction inverted and the flat spectrum of energy.}.
These models do not attempt to describe
all properties of turbulence, but some important ones, such as
the spectrum of energy. They essentially assume
self-similarity, so they don't describe deviations
from self-similarity, known as intermittency. The level
of sophistication of these so-called mean field models is enough
for many applications.

While the spectrum of hydrodynamic incompressible turbulence
is well-established, strong MHD turbulence
is an area of active research (e.g. Montgomery \& Turner (1981),
Shebalin, Matthaeus \& Montgomery (1983), Higdon (1984)).
Numerical research (Cho \& Vishniac 2000, Maron \& Goldreich 2001,
Cho, Lazarian \& Vishniac 2002, Cho \& Lazarian, 2002, 2003, M\"uller, Biskamp \& Grappin 2003)
is roughly\footnote{The particular value of the spectral index and the index of
scale-dependent anisotropy is still an issue of debate (Galtier, Pouquet \&  Mangeney 2005, Boldyrev 2005, 2006, Beresnyak \& Lazarian 2006, Gogoberidze 2007).}
consistent with a particular
model of strong turbulence, called Goldreich-Sridhar model (Goldreich \&
Sridhar, 1995). This mean-field model predicts structural anisotropy
with respect to the magnetic field according to so-called critical balance,
$k_\perp \delta v \sim k_\|v_A$, as well as a $k^{-5/3}$ 
energy spectrum.

The wave dissipation rate has two major applications. First, if it is
directly measured, either in experiment, observation, or
numerical simulation, it can be used as a test of turbulence
itself, and, in certain mean-field models of turbulence, since
the latter rely on the phenomenologically estimated
value of the mean dissipation rate for every particular scale.

The other application is turbulent damping
of linear instabilities. If the instability growth rate is smaller
than the turbulent damping rate, the instability does not develop.
This can have numerous consequences. For instance, it has been customary
to compare the instability rate to the viscous or magnetic dissipation
rates. Because of this, almost any ``large scale'' perturbation
with positive imaginary part of the frequency obtained from linear analysis
has been thought to develop a growing unstable mode. It is possible,
however, that this instability is slow enough to be damped
by the turbulent field.
This way the unstable configuration is stabilized, provided it is not
disrupted by turbulence itself. In the case when turbulence itself is
generated by an instability, the turbulent dissipation
rate will provide a powerful generic mechanism for nonlinear
saturation of the instability. Such a generic saturation mechanism
is highly desirable, since typically nonlinear saturation models
are very specific to the instability in question
and contain quite a number of assumptions.

One example of the application of turbulent damping rates
to astrophysics is the study of the escape of high energy Cosmic Rays
(CRs) from our Galaxy (Yan \& Lazarian 2002, 2004, Farmer \& Goldreich, 2004).
Lazarian and Beresnyak (2006) developed a model in which
the low-energy CR mean free path is effectively reduced in the
presence of compressive motions due to the CR-Alfv\'en
anisotropic instability.

Previous works that considered the turbulent wave dissipation
include Hollweg (1984), Similon \& Sudan (1989),
Kleva \& Drake (1992), Farmer \& Goldreich (2004),
Bian \& Tsiklauri (2007) and others.

In our paper we develop a model of turbulent dissipation
which is purely nonlinear (does not depend on either $Re$ or $Re_m$) and has
self-similar power-law dependencies on the wavenumber, which is
characteristic of turbulence. We also explain why parallel waves
(with wavevector parallel to the field) decay in a different way
than perpendicular waves. We give a general formula for the decay
rate for arbitrarily directed\footnote{We call Alfvenic or pseudo-Alfvenic waves
parallel/perpendicular/arbitrarily {\it directed} instead of {\it propagating},
since these waves always propagate along the local magnetic field, regardless
of the orientation of their wavevector.} waves. The study of the turbulent decay
of parallel directed waves by Farmer \& Goldreich (2004) was motivated
by the fact that such waves have the highest instability grows rate
for CR-plasma instabilities such as the streaming
and the anisotropy instabilities (e.g., Kulsrud 2005).

In what follows we describe the model of turbulent dissipation and
derive the formula for the dissipation rate in \S 2, provide numerical evidence
for turbulent wave dissipation in \S 3, discuss astrophysical implications
in \S 4, mention competing mechanisms for wave damping in \S 5,
discuss previous work in \S 6 and summarize our results in \S 7.

\section{Wave dissipation}

The equations of incompressible ideal MHD,

\begin{equation}
\partial_t{\bf w^+}+({\bf w^-}\cdot\nabla){\bf w^+}=-\nabla P, 
\end{equation}
\begin{equation}
\partial_t{\bf w^-}+({\bf w^+}\cdot\nabla){\bf w^-}=-\nabla P, 
\end{equation}
\begin{equation}
\nabla\cdot{\bf w^+}=\nabla\cdot{\bf w^-}=0, 
\end{equation}

written in terms of Elsasser variables ${\bf w^+=v+b}$ and ${\bf w^-=v-b}$, where
${\bf v}$ is the velocity and ${\bf b}$ is the magnetic field in velocity units
${\bf b=B}/(4\pi \rho)^{1/2}$, bear close resemblance to the Navier-Stokes equations.
This resemblance is somewhat superficial, since ${\bf w^+}$ and ${\bf w^-}$ do not transform
the same way velocity transforms under Galilean transformations. Due to this fact,
there is usually some residual, or local mean magnetic field, that cannot
be excluded by the choice of frame of reference (Kraichnan, 1965).
If we go to sufficiently
small scales the decaying perturbation fields will be much smaller
than the local mean field. This case is called sub-Alfv\'enic turbulence.
The important conservation laws in non-dissipative MHD is the conservation
of the integrals $\int |w^+|^2d^3x$ and $\int |w^-|^2d^3x$, which are the analogs
of conservation of energy in the Navier-Stokes equations.
This allows us to build phenomenology of energy transfer in the spirit of
Kolmogorov (1941) for each of the ${\bf w^+}$ and ${\bf w^-}$ variables with
two different dissipation rates\footnote{In this paper we
consider only the symmetric case in which the kinetic viscosity equals magnetic
diffusivity ($Pr=1$), otherwise some exchange of energy between
Elsasser fields is possible at the dissipation or viscous scale. High Prandtl
number turbulence, which is often relevant for astrophysical processes, was studied
in Schekochihin et al 2002, Lazarian, Vishniac \& Cho 2004, etc.}.
The setup when those dissipation rates are equal is called balanced
turbulence, while the opposite case is called imbalanced
or cross-helical turbulence.

The major difference between the incompressible Navier-Stokes and MHD equations
is that the characteristics of the latter, due to the presence of local
magnetic field, are always wave-like, with the speed of the perturbation being $v_A$.
While it is not possible to have waves in incompressible hydrodynamics,
it is possible to have waves or wave-like perturbations in incompressible
MHD. According to the GS95 model, most perturbations produced
by turbulence cannot be called well-defined waves. But there are other perturbations
generated, for example, by instabilities, which, as we will demonstrate in this paper,
have frequency much larger than decay rate and, therefore, can be called waves.

This paper is focused on the study of the decay of a small perturbation
in stationary forced turbulence. Unlike decaying turbulence, where
the energy decays self-similarly, or as a power-law with time, the decay
of a small perturbation in stationary turbulence will be exponential.
This statement is based on the general concept of the flow of energy
through scales, which underlies all Kolmogorov-type models.
Indeed, in Kolmogorov models, the decay rate is some function
of the perturbation strength and the length scale: $\gamma_{\rm turb}(\delta v, \lambda)$.
For example, in strong hydrodynamical turbulence, $\gamma_{\rm turb}=\delta v/\lambda$.
Since the wave amplitude is small, it can not significantly affect $\delta v$. Therefore
the dissipation rate will be constant, meaning that the wave will decay exponentially.

In this paper we only consider the Alfv\'en and pseudo-Alfv\'en waves
that exist in incompressible turbulence. The study of the decay
of all three wave modes in generic compressible turbulence
is a task that will be addressed elsewhere.

Now we will provide a phenomenological
description of wave decay and the formula for the decay rate
for arbitrarily directed waves. To highlight the general problem
we first consider parallel waves.
According to the second-order perturbation theory, such waves
are not cascaded (e.g., Galtier et al 2002). Does
this property necessarily apply to strong turbulence?
Farmer \& Goldreich (2004) argued that this is not the case,
addressing the issue of the wave propagation parallel to the local
magnetic field when the effective perpendicular wavenumber
is obtained by so-called field wandering.
Let us imagine a plane wave with very large perpendicular
correlation length, corresponding to $k_\perp=0$ and finite
parallel correlation length $\Lambda$, corresponding
to $k_\|=1/\Lambda$. This wave, moving a distance of $\Lambda$,
will get disrupted by field wandering, as various pieces of this
wave propagate along its local field direction, which is slightly
different along the transverse spread of the wave. Note that only
field-wandering from the counter-propagating waves will contribute
to this process, as the co-propagating waves will move exactly the
same distance, being stationary in the frame of the wave.
The ``effective'' perpendicular wavenumber $k_{\perp, {\rm eff}}$ is determined by the condition
$k_{\perp, {\rm eff}}=(\delta B/B) k_\|$, where we use $\delta B$ of counter-wave
on scale $1/k_{\perp, {\rm eff}}$.

The arguments above can be generalized for the case of an arbitrarily
directed wave in the following way.
While the wave is being decorrelated by turbulence, it obtains
a new ``effective'' $k_{\perp, {\rm eff}}$ from the vector sum of the original $k_\perp$ and the one
produced by turbulent field wandering. The effective $k_{\perp, {\rm eff}}$ in the
RMS sense will be the square mean of the two.
If the original wave has a
pitch angle $\theta$ with respect to the field and a
wavenumber magnitude of $k$, its effective perpendicular wavenumber
will be determined by the equation

\begin{equation}
k_{\perp, {\rm eff}}=k(\sin^2\theta+(\delta B/B)^2\cos^2\theta)^{1/2}, 
\end{equation}

or, if we assume the GS95 scaling $\delta B/B \sim k_\perp^{-1/3}$,

\begin{equation}
k_{\perp, {\rm eff}}=k(\sin^2\theta+(k_{\perp, {\rm eff}} L)^{-2/3}\cos^2\theta)^{1/2}. 
\end{equation}

According to the arguments above,
we should include only the part of $\delta B/B$ which corresponds
to the counter-wave, since co-propagating perturbations do not influence
the structure of the wave. This is due to the exact solution
of the incompressible MHD which involves only one wave species of
arbitrary amplitude. In eq. (5), for the sake of simplicity, we omitted
the coefficient $1/2$ before
$\delta B/B$ which would indicate that only half of the perturbations
contribute to the decorrelation effect.

It is easy to verify from eq. (5), that for perpendicular modes
$k_{\perp, {\rm eff}}=k\sin\theta=k$ (as if field wandering is unimportant),
while for parallel modes, $k_{\perp, {\rm eff}}\sim k^{3/4}$ (as in Farmer \& Goldreich 2004).

The decay rate is determined by the $k_{\perp, {\rm eff}}$
we calculated from (5):
\begin{equation}
\gamma=v_A k_{\perp, {\rm eff}}^{2/3}L^{-1/3}. 
\end{equation}
Eq. (6) deserves a more detailed explanation. While cascading
of perturbations in {\it weak turbulence} have resonant
character, that is, they require counter-waves with equal
and oppositely directed ${\bf k_\|}$ (e.g., Ng \& Bhattachargee 1996,
Goldreich \& Sridhar 1997, Galtier et al 2002),
{\it strong turbulence} involves nonlinear
perturbations where strict resonance condition is not required.
This can be demonstrated if we consider a cascaded perturbation
as a passively advected vector field. This approach is valid
as long as the amplitude of this sample or test perturbation
is small enough to avoid any back-reaction to fully developed,
balanced strong turbulence. Now, according to critical balance of GS95,
the transverse structure of this passively advected field
is strongly perturbed by counter-waves with the same $k_\perp$
after traveling a distance of $1/k_\|=k_\perp^{-2/3}L^{1/3}$.
On the other hand, according to GS95, this is the correlation
length of the counter wavepacket, so the ``next''
perturbation will be applied to our test field incoherently.
Therefore, these perturbations of the transverse structure
of the test field are irreversible and can be called cascading
in perpendicular direction. Our main point here is that
the lateral structure of the {\it test field} is irrelevant
for its cascading (unless the lateral structure produces
the transverse structure by field wandering), as long as cascading
happens in strong GS95 turbulence\footnote{This way the test-field
approach is different from the dynamic model of GS95 itself,
where the lateral structure of the counter-waves is essential for
back-reaction on co-waves and creating its own critical balance.}.

The formulae (5) and (6) give the general dependence of the
decay rate $\gamma$ on arbitrarily directed (oblique) waves
with wavevector $k$ and pitch-angle $\theta$.
This dependence is, in general, not a power-law of $k$.
There are two limiting cases of
the perpendicular directed and the parallel directed waves which give
power-laws, namely

\begin{equation}
\gamma_\perp=v_A k^{2/3} L^{-1/3}, \ \ \  \gamma_\|=v_A k^{1/2} L^{-1/2}. 
\end{equation}

Fig. 1 shows the dependence of the decay rate $\gamma$ on the wavevector $k$
for a variety of pitch angles $\theta$ with respect to the magnetic field.

\begin{figure}
\plotone{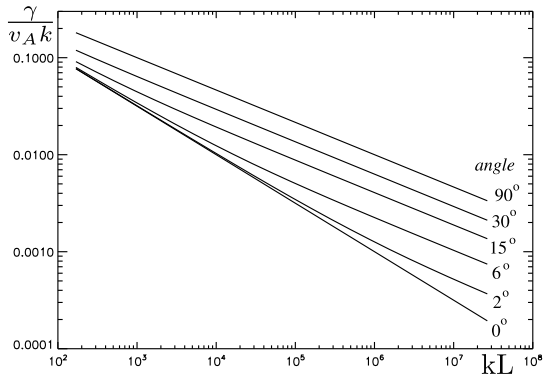}
  \caption{The relative decay rate versus wavenumber for different angles
   as obtained from Eqs. 5 and 6.}
\label{gamma_theor}
\end{figure}

\section{Numerical Results}
We have conducted a series of 71 incompressible three-dimensional $320^3$
MHD simulations with stochastic turbulent driving. We used a pseudo-spectral
code to solve Eqs. 1-3 and introduced linear dissipation as an energy sink.
We used relatively low-order (6th order) hyperdiffusion to extend the
inertial range while avoiding a strong bottleneck effect. The turbulent
driving supplied energy between $k=2$ and $3.5$ while most of the
dissipation happened around $k=70$. We regarded the interval $k=5$ to $50$
as an inertial interval of turbulence. Simulations had an
external magnetic field and the driving was tuned to produce
strong turbulence which was nearly isotropic around the driving scale.
More details of the code and the driving can be found in
Cho and Vishniac (2000). 

In this series one of the simulations was without wave driving,
and all others were driven with a plane wave with a particular
wave vector ${\bf k}$. The turbulent driving and the initial conditions
in all series were exactly the same. The initial conditions were taken
from another turbulent simulation with driving that ran for around 20
Alfv\'en times. We waited for about 0.3 Alfv\'en times for the wave
to reach saturation. We chose the wave driving amplitude so that
the saturated wave does not incur any significant back-reaction
to turbulence. After this the wave driving was switched off but
the turbulent driving continued just like in the simulation
without wave driving.

As expected, after the wave reached saturation,
it was no longer plane-parallel and
its energy distribution in k-space was affected by both field line
wandering and strong turbulent cascading.
It is more appropriate to say that the simulation box was filled with a collection
of wave packets which had the dispersion in the direction of ${\bf k}$ {\it and}
the dispersion in $k_\|$ and $k_\perp$ that comes from turbulent decorrelation
and the corresponding uncertainty relation. In Fig. 2 we presented a visual
representation of the distribution of wave energy that was initially driven
as the parallel wave with $k_\|=17$ (in cube size units) and reached saturation.

\begin{figure}
\plotone{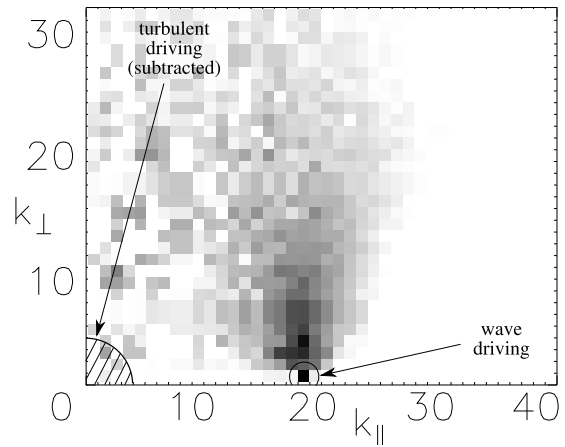}
  \caption{Decorrelation of the parallel-propagating wave by turbulent
   field wandering. Darker areas indicate larger wave energy. The wave
   is decorrelated and cascaded primarily in the direction perpendicular
   to the magnetic field.}
\label{decorrel}
\end{figure}

The wave energy was obtained by subtracting the results of the simulation with
and without wave driving. We summed the result of subtraction
only over the relevant regions in $k$-space, to reduce errors. 
We note again, that driving, although being turbulent-like, was predetermined,
and within our short simulation times (about 0.6 Alfv\'en times) chaos
in motions on large scales was not yet able to develop. This was confirmed by the fact
that the result of subtraction was mostly positive and significantly differed 
from zero only around the region where we pumped the wave. One notable exception
from this rule was the velocity at k=0, or the mean velocity. It increased,
which is interpreted as a transfer of momentum from the wave to the fluid. 
We have carried out a total of 70 $320^3$ simulations with different values
of the wave ${\bf k}$ vector which have covered the inertial range
of our turbulent simulation from $k=10$ to $k=40$ with different angles
with respect to the mean magnetic field.

After the wave driving was switched off, the energy of the wave started to decay exponentially
(see Fig. 3). We measured the decay rate by fitting a decay curve.
Fig. 4 shows the dependence of the decay rates on the $k$ value and on the angle
of original inclination of wave to the magnetic field (the angle $\theta$ between ${\bf k}$ and ${\bf B}$).
As we see, for waves parallel to the external field the decay rate follows
a $k^{1/2}$ law (see also Farmer \& Goldreich 2004).
The theoretical curves for the different inclinations that were fitted into
numerical values of $\gamma$, unexpectedly, have slightly different
values for the outer scale $L$. We believe that this could be either due
to the fact that the wavelengths that we considered were relatively close
to the driving scale, or the fact that we simulated trans-Alfv\'enic
turbulence.

\begin{figure}
\plotone{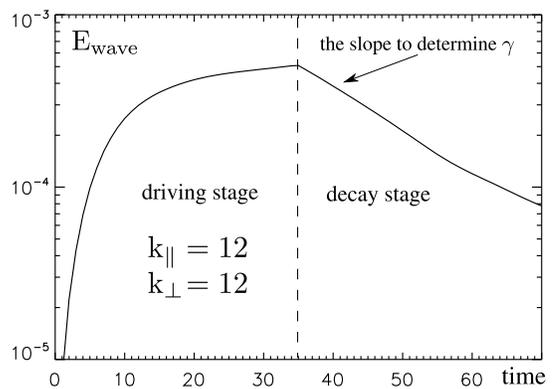}
  \caption{The evolution of the oblique wave energy during the driving and decay stages.
           The wave driving was at $(k_\|, k_\perp)=(12,12)$ in code units, approximately
           $(4, 4) L^{-1}$.}
\label{wave_decay}
\end{figure}

\begin{figure}
\epsscale{.6}
\plotone{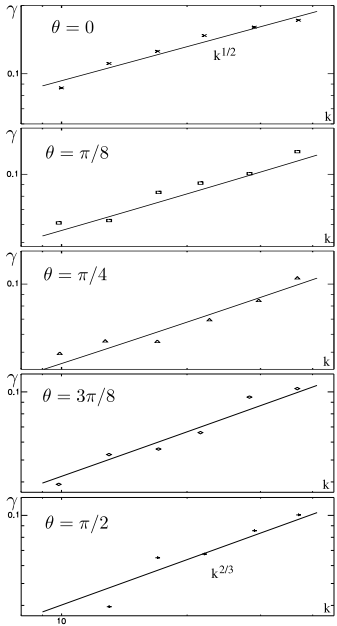}
\epsscale{1}
  \caption{The dependence of the decay rate on the magnitude of the wavevector for
different angles between ${\bf k}$ and ${\bf B_0}$ in a direct 3D numerical simulation (points),
fitted with curves obtained from eqs. 5 and 6 (lines).}
\label{gamma_simul}
\end{figure}

\section{Applications of Turbulent Damping}
The damping of Alfven waves with ${\bf k}$ parallel to the
magnetic field was invoked by Farmer \& Goldreich (2004) in the context
of the damping of the streaming instability by ambient turbulence
(first mentioned in Yan \& Lazarian 2002). They estimated that with the
modest level of turbulence in our Galaxy, the streaming instability
will be suppressed for cosmic rays with energies higher than
about 100 GeV; therefore such cosmic rays can freely propagate
along the magnetic field lines.
Lazarian \& Beresnyak (2006) have used turbulent damping
as one of the mechanisms that limits the CR anisotropy instability
which occurs in a magnetized fluid in
the presence of compressible turbulence. This instability greatly
decreases the CR's mean free path. However, due to turbulent
damping, there is a high energy limit of around 1000 GeV
(for our Galaxy) for this mechanism. Particles with energy
higher than this limit will be unaffected by the instability
and scattered primarily through other mechanisms, such as direct
interaction with MHD modes (Chandran 2000, Yan \& Lazarian 2002, 2004, 2007). 

The influence of the turbulent decay to the development of
instabilities can lead to the various interesting astrophysical
effects. Let us consider one example.  In a recent paper, Everett et al
(2007) (henceforth E07) considered a model of launching the Galactic
wind by CR pressure (see also Breitschwerdt et al, 1991). They
argued that CR pressure is necessary to launch the wind, since
thermal pressure is too small.  One of the critical assumptions of
this model is that the streaming instability operates effectively and
allows CRs to exchange momentum with the plasma.  Let us now
assume that the streaming instability is damped by turbulence.
According to Farmer \& Goldreich (2004), if the amplitude of turbulence is
similar to that of our Galaxy, the damping of the streaming instability
will affect CRs with energies higher than 100 GeV, making them
effectively disconnected from the background plasma and able to
escape.  If most of the energy (and the momentum) of CRs are in the lowest
energy CRs, as E07 originally assumed, this would present no problem
for their model.  It is believed, that our Galaxy, on average, has a
relatively steep spectrum of CRs, between the slopes of $-2.6$ and
$-3.1$, justifying this assumption. However, the winds are most
likely to be launched from the sites of intensive
CR production, such as regions with many supernova shocks. Some modern
theories of shock acceleration with precursor favor a rather shallow
spectrum of the accelerated particles, such as shallower than $-2$ (Malkov
\& Drury 2001) \footnote{The average galactic value of $-2.6$ is then
  obtained by a variety of mechanisms such as re-acceleration,
  escape, etc.}. CRs with such a spectrum have most of their energy and momentum
in higher energy particles that can freely escape because of damping
of the streaming instability and, therefore, CR pressure is too weak to launch the
wind. The existence of the Galactic wind would therefore suggest that it is
always being launched by a CR population with a steep spectrum, or that these winds
have very low level of turbulence. Since we cannot directly observe
the spectrum of the CRs near acceleration sites such information about the slope
being shallower or steeper than $-2$ can be used to limit theories of
CR acceleration in shocks.

The other application of energy dissipation of Alfv\'en
waves is the problem of coronal heating. It has been long known
that due to the very small values of viscous and magnetic dissipation
coefficients, ohmic and viscous heating are unlikely to dissipate
Alfv\'en waves significantly. On the other hand, we know that
a relatively small fraction of energy of waves actually escapes the
coronal region. Turbulent dissipation can deal with this problem,
since it provides much higher dissipation rates than ohmic
and viscous heating, and, unlike the latter, does not depend
on the Reynolds numbers $Re$ and $Re_m$. Parker (1991) noted,
however, that in the corona, the stochastic component of the field
mostly consists of waves propagating in one direction,
outwards from the Sun. In our terminology this is a strongly imbalanced
case. While there is certainly a strong imbalance close to the photosphere
near the source of the waves, the situation in the upper
solar corona could be closer to balanced turbulence due to various mechanisms
that allow waves to reflect back, such as the geometry of the guiding
magnetic field, the parametric instability, etc. Despite coronal heating still
being a poorly understood process there are few things that we
would like to note with regards to our model. First of all, our model
does not consider dissipation in imbalanced turbulence, due
to the fact that models of strong imbalanced turbulence
are being developed (see Beresnyak \& Lazarian 2007 and references therein, see Discussion).
Secondly, we only considered waves with small amplitudes
that do not produce strong back-reaction to turbulence.
We hope, however, that the present work can provide
insight into a more complicated picture that will include
imbalanced turbulence, reflection of waves, etc.

There is also the problem of launching winds, such as a stellar
wind, by momentum from waves. It is similar to the previous
problem, in that we need mechanisms of wave reflection, in order
to generate turbulence, otherwise waves will freely escape.
It is well known that the solar wind is strongly imbalanced
close to the Sun, but becomes less imbalanced at larger distances, almost reaching
equipartition between waves near the Earth's orbit.
Roberts et al (1987) suggested that this was due to the
dominance of the kinetic energy on the outer scale.

\section{Competing Mechanisms of Damping}

While the ohmic and viscous dissipation rates decrease very rapidly
with increasing wavelength, as $k^2$, there are a number
of mechanisms of Alfv\'en wave dissipation that, in principle,
can compete with turbulent dissipation. The dissipation mechanisms
can be subdivided into linear and nonlinear mechanisms. We call it
a linear mechanism when the decay rate does not depend on the amplitude
of the wave. Such mechanisms include viscous damping, turbulent damping (even
though the turbulence itself is a nonlinear process), collisionless
(Landau) damping, and the resonance damping. Nonlinear damping
mechanisms have a decay rate that depends on the wave amplitude.
These include nonlinear Landau damping, wave steepening, etc.
 
Nonlinear Landau damping (e.g., Kulsrud 2005) requires
two waves with close frequencies and a population
of ions that is in resonance with the beat wave of those
two waves. While, generically, the decrement of this damping
is proportional to the frequency, it has a nonlinear
$(\delta B/B_0)^2$ factor that limits damping to high
amplitude waves. Note that the turbulent damping mechanism that
we consider in this paper has a substantial limitation in that the
wave amplitude has to be small enough to not back-react on the
turbulence itself. Thus, nonlinear damping mechanisms and the
linear (with respect to the wave amplitude) turbulent
damping are, in a sense, complimentary to each other (see fig. 5).
Another nonlinear mechanism of damping
is wave steepening (e.g., Cohen \& Kulsrud, 1974).
This mechanism is also supposed to have $\gamma \sim kv_A(\delta B/B_0)^2$.

\begin{figure}
\plotone{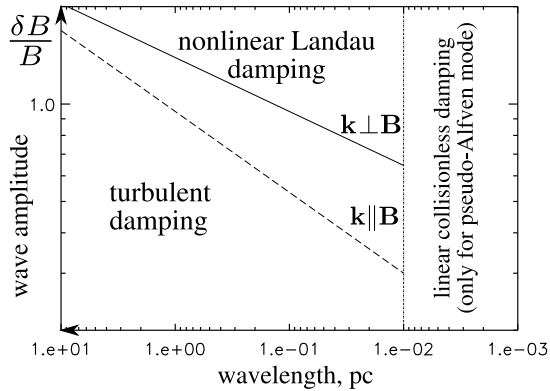}
  \caption{Dominating dissipation mechanisms for various wave lengths and
wave amplitudes in the ICM or hot ISM. Linear collisionless (Landau) damping acts only on the
pseudo-Alfven mode, while nonlinear Landau damping and turbulent damping
operate on both modes.}
\label{nl_landau}
\end{figure}

The prominent linear mechanism of damping is so-called resonance
damping (e.g., Hollweg, 1984). Unlike previously mentioned
mechanisms that are considered in a homogeneous medium and usually
work equally well in any geometry, this linear damping mechanism appears
when, due to the specific boundary conditions or inhomegeneity of the flow,
the so-called Alfvenic resonance appears. This resonance corresponds
to the flow speed being equal to the phase speed of the Alfvenic
wave. The resonance is typically the $1/x$ type divergence for the magnetic
field and velocity. Notably, the dissipation provided by this resonance
does not depend on viscosity or magnetic diffusivity (Kappraff \& Tataronis 1977).
The dissipation rate can be calculated from the ideal MHD equations with $\nu=\eta=0$
using contour integrals (e.g., Livshitz \& Pitaevskii 1981,
Pariev \& Istomin 1996). The Alfven resonance of the flow has been proposed as a prominent
candidate for an in-situ acceleration mechanism in jets (Beresnyak, Istomin \& Pariev 2003).
Resonance damping usually provides damping rates that are a fraction
of the wave frequency.
It is considered to be important when the wave length is of order
the scale at which boundary conditions are set. In jets we expect
turbulent damping to be more important than resonance damping
(if the latter is present) for wavelengths that are much smaller
than the jet's diameter, assuming that for these wavelengths the medium
can be treated as homogeneous.

\section{Discussion}
This paper deals with wave dissipation in a {\it balanced}
MHD turbulence. The more general imbalanced case was not considered
for a variety of reasons. First, the theory of strong imbalanced MHD
turbulence is still being developed (Lithwick, Goldreich \& Sridhar 2007,
Beresnyak \& Lazarian 2007). On the other hand, if we adopt the model
of imbalanced turbulence presented in Beresnyak \& Lazarian 2007 (furthermore BL07)
a number of questions arise. Most importantly, we will be unable to reproduce the
statement of \S 2 that the cascading of the wave is irrelevant to its lateral
structure. This is due to the fact that neither the strong nor the weak wave from BL07
has a critical balance with itself (i.e. $k_\|v_A=k_\perp\delta v$ does not
hold if all quantities refer to the same wave, instead BL07 provides a different
expression for the critical balance of each wave). Therefore there is
no guarantee that we can use the strong cascading formula, at least when the
wave has a large $k_\|$.

An attempt to include irregularities of magnetic field lines due
to turbulence to increase predicted dissipation in solar corona was
made in Silimon \& Sudan (1989).  Their model relies on a description
of turbulent magnetic field lines as divergent with a specific
Lyapunov constant, or exponentiation rate $\lambda$. We argue that
this is not a proper description of turbulent fields in developed
turbulence because such fields are approximately self-similar and do not have any
designated scale in its inertial interval. In developed turbulence
every scale has its own exponentiation rate. Silimon \& Sudan (1989)
give a formula for dissipation length which is inversely proportional
to the Lyapunov constant (which is not specified, but estimated) and
only weakly (logarithmically) proportional to the wavenumber.  It also
depends on magnetic Reynolds number.
This paper, in contrast, advocates purely nonlinear dissipation
rates that do not depend on $Re_m$ or $Re$.
Parker (1991) criticized Silimon \& Sudan's approach of using stochastic field
lines to increase dissipation by noting that
the stochastic component of the field is itself part of the
waves propagating outwards along the field. We commented on this
controversy previously by noting that one has to have a mechanism
to reflect the waves back, otherwise they escape freely.

Kleva \& Drake (1992) have studied nonlinear dissipation of waves
in the presence of a stochastic field on an outer scale by numerical
methods. They confirmed that the dissipation of large scale waves
does not depend on dissipation coefficients, while for small scale
waves it approached the viscous and resistive limit. However, their
stochastic field was not turbulent, but rather a predetermined field
on the outer scale. Also, the damping rate they measured did not depend
on wavevector as a power-law.

Bian \& Tsiklauri (2007) have considered mixing of Alfv\'en wavepackets
in chaotic magnetic fields. They obtained an analytical
expression for the advection-diffusion of wavepackets using
a WKB approximation and claimed that the stationary wave energy
spectrum is $k^{-1}$. We feel that the WKB method is not the appropriate
tool to derive turbulent dissipation, as it requires
that the wave should have wavelengths that are much smaller than the
underlying perturbations, while in turbulence the most effective
nonlinear dissipation comes from scales comparable with the
wavelength.

Hollweg (1984) considered dissipation of Alfv\'en waves
in coronal loops, primarily due to resonance mechanism,
but also speculated on turbulent heating. He took the Kolmogorov
dissipation rate (which is a decay that happens on
a kinetic timescale $\lambda/v$) with the perturbation
length scale as wavelength and the RMS wave velocity
as perturbation velocity. This way he obtained
dissipation which is compatible with heating in
coronal loops as well as coronal holes. We have to note,
however, that propagating waves do not necessarily
generate turbulence and even if there is a flux
of waves in both directions, there are certain
requirements for MHD turbulence to be strong
turbulence and have the Kolmogorov dissipation rate.
For example, in the context of coronal loops, with a strong
guide field, it is easy to imagine a situation where
turbulence is not strong on the outer scale, therefore
it has dissipation rate lower than Kolmogorov.

\section{Summary}

In this paper we demonstrated the following:

1. Parallel and perpendicular directed waves
are dissipated in different ways. Perpendicular
waves are cascaded naturally by counter-waves,
while parallel waves are first decorrelated
by the field wandering and then cascaded.

2. In the inertial interval of sub-Alfv\'enic turbulence,
parallel waves are damped more slowly than perpendicular
waves with the same wavenumber. The decay rate for
the parallel wave is much smaller than its frequency,
it is not eddy but a well-defined wave.

3. In a homogeneous medium with open boundaries turbulent
dissipation is more effective than the other dissipation
mechanisms, provided that the wave amplitude is sufficiently
small and the wavelength is in the inertial interval
of turbulence.

\acknowledgments
We thank Jungyeon Cho for providing the initial version of the code
and for some ideas on wave propagation in turbulence.
AB thanks John Everett and Vladimir Pariev for fruitful discussions. 
AB thanks IceCube project for support of his research.
AL acknowledges the  NSF grant AST-0307869 and the support from
the Center for Magnetic Self-Organization in Laboratory and Astrophysical
Plasmas.

\end{document}